\documentclass[aip,graphic]{revtex4-1}
\usepackage{amssymb, siunitx, graphicx}
\draft 

\begin{document}


\title{Comparison of intermediate-range order in GeO$_2$ glass: molecular dynamics using machine-learning interatomic potential vs.\ reverse Monte Carlo fitting to experimental data} 



\author{Kenta Matsutani}
\affiliation{Graduate School of Science and Engineering, Yamagata University,
1-4-12 Kojirakawa, Yamagata, 990-8560, Yamagata, Japan}
\author{Shusuke Kasamatsu}
\affiliation{Faculty of Science, Yamagata University, 1-4-12 Kojirakawa, Yamagata, 990-8560, Yamagata, Japan}

\author{Takeshi Usuki}
\affiliation{Faculty of Science, Yamagata University, 1-4-12 Kojirakawa, Yamagata, 990-8560, Yamagata, Japan}



\date{\today}

\begin{abstract}
The short and intermediate-range order in GeO$_2$ glass are investigated by molecular dynamics using machine-learning interatomic potential trained on ab initio calculation data and compared with reverse Monte Carlo fitting of neutron diffraction data. To characterize the structural differences in each model, the total/partial structure factors, coordination number, ring size and shape distributions, and persistent homology analysis were performed. These results show that although the two approaches yield similar two-body correlations, they can lead to three-dimensional models with different short and intermediate-range ordering.
A clear difference was observed especially in the ring distributions; RMC models exhibit a broad distribution in the ring size distribution, while neural network potential molecular dynamics yield much narrower ring distributions. This confirms that the density functional approximation in the ab initio calculations determines the preferred network assembly more strictly than RMC with simple coordination constraints even when using multiple diffraction data.
\end{abstract}


\maketitle 

\section{Introduction}
Network--forming amorphous materials have been actively studied for their structural characterization and understanding because of their excellent mechanical, optical, and transport properties \cite{angell1995, thorpe1983}.
The atomic arrangements in amorphous materials like silica, germania, zinc dichloride, B$_2$O$_3$, and others show three distinct length scales, which are observed in X-ray or neutron diffraction experiments measuring the structure factor\cite{zeidler2016a, salmon2019, onodera2020, salmon2023}. 
The third maximum peak implies the nearest neighbor distance, and the second peak, the principal peak, reflects the size of the local structural motifs such as SiO$_2$ tetrahedra in silica glass. The first peak, known as the first sharp diffraction peak (FSDP), which correlates with the intermediate-range order (IRO), is proposed to represent the arrangement of these structural motifs \cite{salmon2023}.
Numerous interpretations of the structural origin of FSDP have been proposed, ranging from quasicrystalline and ring-like structures similar to those found in crystals \cite{gaskell1996} to layered structures \cite{lin1984} and cavities \cite{massobrio2001}. The difficulty in interpreting FSDP arises from the fact that diffraction experiments provide only one-dimensional reciprocal space structure information. 
Therefore, computer simulation methods such as reverse Monte Carlo (RMC) \cite{keen1990, evrard2005, morita2009} and molecular dynamics \cite{shanavas2006, marrocchelli2009, zhu2009, li2009, wezka2012, sorensen2020, erhard2022, kobayashi2023} have been used to create three-dimensional structural models to obtain information beyond these limitations. 

In this work, we focus on GeO$_2$ glass, which is known as a prototypical oxide network-forming material. The local structural motifs are known as tetrahedral units centered on a Ge atom coordinating 4 oxygen atoms, and these units form a network through the oxygen atom to which the two Ge atoms are coordinated \cite{micoulaut2006a}.
Structural analysis based on simulations and experiments have been actively conducted for this system \cite{kohara2005, shanavas2006, zhu2009, li2009, marrocchelli2009, matsutani2022a}; 
a brief review and RMC analysis results compared with molecular dynamics and experimental results are given in our previous work \cite{matsutani2022a}. 
However, structural details, especially of IRO, are still under debate. This is partly due to the fact that diffraction experiments only provide one-dimensional information as noted above; as a result, obtaining reasonable structures using the RMC method depends crucially on the initial structure and physical constraints which need to be known beforehand.
On the other hand, melt-quench simulations using classical molecular dynamics can often provide a reasonable glass structure, but the agreement of the calculated structure factor with the experimental one is often unsatisfactory.
One way to overcome these problems is to combine molecular dynamics and RMC analysis.
Previous works have often performed melt-quench simulations of glass formation using classical molecular dynamics, whose resulting structure is used as the initial structure for subsequent RMC modeling \cite{gereben2012, onodera2019, shiga2023}. 
A downside of this approach is that the final results depend on the RMC method, and the obtained structures may be energetically unfavorable and unphysical. 

The above issues may be overcome if an accurate force field was available that can reproduce the structure factors. Towards this end, ab initio molecular dynamics (AIMD) calculates forces acting on atoms on-the-fly based on quantum mechanics principles, but as a result, the calculations are exceedingly computationally demanding; this restricts the simulation timescale to picosecond order and the number of atoms to only a few hundred. Therefore, this technique does not allow for sufficient structural relaxation of the glass, and the simulation cell size is too small (typically less than 1000 \r{A}$^3$) to observe IRO (the correlation length of IRO is proposed to be \SIrange{10}{20}{\angstrom} \cite{elliott1991, salmon2019, phillips1980, salmon1994}).

In the last few decades, a possible solution to this issue has emerged in the form of machine learning potentials, which have been rapidly developed and successfully applied to a wide variety of materials \cite{behler2017, westermayr2020, watanabe2020}. Unlike classical interatomic potentials, which have analytical forms based on physics and chemistry viewpoints, machine learning potentials rely on no such physical/chemical intuition and are comprised of neural networks \cite{behler2007, artrith2017}, Gaussian process \cite{bartok2015, jinnouchi2019}, tensor expansion \cite{shapeev2016}, etc. These functional forms are fit to (or, in machine-learning terminology, trained on) ab initio energy and force data. This method offers tremendous computational cost advantages over ab initio molecular dynamics while preserving the accuracy of the reference datasets. It is applicable to diverse systems, such as phase change materials, that require an appropriate description of bond formation and breakage \cite{lee2020}.
We have taken this route in this work and constructed neural network potentials for GeO$_2$ glass, evaluated its accuracy, and investigated the resulting glass structure. 

Even if we are able to obtain reliable three-dimensional structural models using the above-discussed method, a challenging task remains in interpreting the intermediate range order in glass systems. In the past, the bond angle, dihedral angle distribution, and ring size distribution \cite{king1967a, guttman1990, goetzke1991, leroux2010a} have been used to characterize the intermediate-range order in glass. 
The former methods represent 3- or 4-body correlations and can describe correlations beyond the nearest neighbor. The latter method can capture more extended structural features based on chemical bonds. While conventional ring analysis achieves some success, it is based on size distribution and neglects ring shape and length scale information. 
Some researchers have proposed the calculation method and definition of ring shape to overcome these limitations \cite{salmon2023, shiga2023, kobayashi2023}.
In recent years, another purely mathematical method called persistent homology, which was developed in the 2000s as a tool for analyzing topological features in multiscale data, has been applied to atomistic structures.
Persistent homology is one of the fundamental concepts of topological data analysis (TDA) and is widely used in the investigation of glass structures as a tool for extracting structural features beyond local structural motifs. A review of the concepts and applications can be found in refs.~\cite{hiraoka2016, sorensen2020, obayashi2022}. We apply these recent developments to investigate the IRO of the glass structure obtained using NNP-MD and RMC modeling.

This paper is organized as follows. The main methods used, such as molecular dynamics and analysis details, are described in the Methods section. The Results and Discussion section focuses on the dependence of local and intermediate order in GeO$_2$ glass on the neural network potentials molecular dynamics (NNP--MD) results compared to RMC analysis. The conclusion section summarizes this work.

\section{Methodology}
\subsection{Training data collection and construction of force field}
All AIMD calculations were conducted by Vienna Ab initio Simulation Package \cite{kresse1996} to generate datasets for NNP training. The cutoff energy of the plane wave basis set was \SI{289.8}{\electronvolt}, and the \textit{k}-point sampling used only the $\Gamma$ point. The self-consistent-field energies were converged to within \SI{e-5}{\electronvolt}. The projector augmented wave method was used to describe electron--ion interactions and the standard density functional approximation Perdew, Burke, and Ernzerhof (GGA--PBE) \cite{perdew1996}. A melt--quench process was simulated in AIMD using GGA--PBE with a 96--atom model to obtain glass structure from crystal. The glass and crystal structure under isotropic pressure were also simulated. A small amount of data with different stoichiometries (Ge$_3$O and GeO$_3$) are also calculated to prevent phase separation in the NNP-MD \cite{lee2020}. The training dataset was then extracted from those simulation trajectories, which finally contained 1870 structures; the details of the dataset are described in Table \ref{tab:train}. Then, we recalculated the extracted 1870 data with a cutoff energy of \SI{500}{eV} and a k-point sampling of 2 2 2 to obtain high-quality energy and interatomic force data for training the NNP. \footnote{It is noted that the NNP trained on the data with lower cutoff and k-point mesh were not reliable when running long-time simulations and resulted in unphysical phase separation} The self-consistent-field energies were converged to within \SI{e-6}{\electronvolt}.
All molecular dynamics calculations are performed using Nos\'e-Hoover thermostat under \textit{NVT} conditions.
To confirm whether the learning of NNP had progressed properly, we performed an additional AIMD of the quenching process and collected 200 structures as a test data set. The calculation parameters were the same as those described above.

We employed SIMPLE-NN code \cite{lee2019} for the generation of our neural network potentials from the above dataset, which is separated into training and validation data with a ratio of 9:1.
The structure of neural networks is 70-30-30-1. We employed atom--centered symmetry functions \cite{behler2007, behler2015, behler2017} as the descriptors, which consisted of 16 radial and 54 angular components (\textit{G2} and \textit{G4}), and the cutoff radius is \SI{6.0}{\angstrom}. The specific parameters for each of the symmetry functions are the same as those in Ref. \onlinecite{artrith2016, lee2020}. The loss function is defined as the sum of RMSE for total energy and force. The network was optimized by Adaptive Momentum Estimation \cite{kingma2014} (Adam) with 0.001 learning rate, and the batch size was set to 10.

\begin{table}[hbtp]
    \caption{Summary of the training dataset. The first column describes the molecular dynamics calculation with respect to the trajectory, and the second column shows how the structures were extracted from those trajectories. The third describes the number of structures.}
    \label{tab:train}
    \centering
    \scalebox{1}{
    \begin{tabular}{c|c|c}
        \hline
         & Extracted structures from MD trajectory & Number of structures \\
        \hline \hline
        crystal (300 to 3000 K) & Every 1 ps from 100 ps MD & 100\\
        liquid (3000 K) & Every 50 fs from 25 ps MD & 500 \\
        quench (3000 to 1000 K) & Every 1 ps from 100 ps MD & 100 \\
        quench (1000 to 300 K) & Every 350 fs from 35 ps MD & 100 \\
        glass (\SIrange{0.0682}{0.0796}{\angstrom^{-3}}) & Every 20 fs from 1 ps MD & 50\\
        glass (\SIrange{0.0796}{0.0936}{\angstrom^{-3}})&  Every 20 fs from 1 ps MD& 50\\
        glass (\SIrange{0.0936}{0.1073}{\angstrom^{-3}}) & Every 20 fs from 1 ps MD& 50\\
        glass (\SIrange{0.0682}{0.1073}{\angstrom^{-3}}) & Every 10 fs from 500 fs MD& 50\\
        glass (\SIrange{0.0682}{0.0448}{\angstrom^{-3}}) & Every 20 fs from 1 ps MD & 50\\
        crystal (\SIrange{0.0682}{0.0796}{\angstrom^{-3}}) & Every 50 fs from 10 ps MD& 200\\
        crystal (\SIrange{0.0796}{0.0936}{\angstrom^{-3}}) & Every 50 fs from 10 ps MD& 200\\
        crystal (\SIrange{0.0936}{0.1073}{\angstrom^{-3}}) & Every 50 fs from 10 ps MD& 200\\
        crystal (\SIrange{0.0682}{0.1073}{\angstrom^{-3}}) & Every 5 fs from 1 ps MD & 200\\
        liquid Ge$_3$O (\SI{0.0682}{\angstrom^{-3}}) & Every 50 fs from 500 fs MD & 10\\
        liquid GeO$_3$ (\SI{0.0682}{\angstrom^{-3}}) & Every 50 fs from 500 fs MD & 10\\
        \hline
    \end{tabular}
    }
\end{table}

\subsection{Modeling procedure}
NNP-MD simulations of glassy GeO$_2$ were performed using the Large-scale Atomic/Molecular Massively Parallel Simulator (LAMMPS \cite{plimpton1995, thompson2022}). Firstly, the structure was extracted from the trajectory of an AIMD simulation of liquid GeO$_2$ consisting of 120 atoms, and each side of the structure was tripled to create a system of 3240 atoms. Then, the extended model served as the initial structure and was maintained at \SI{2500}{\kelvin}, using the trained NNP with Nos\'e-Hoover thermostat under \textit{NVT} conditions for \SI{1}{\nano\second}. Subsequently, all models were quenched to \SI{300}{\kelvin} over a period of \SI{1}{\nano\second} (cooling rate \SI{2.3}{\kelvin / \pico\second}). The obtained glass structures were held at room temperature and zero pressure for \SI{1}{\nano\second}.
The above procedure was repeated independently three times to obtain three glass structures for evaluating the variation of the calculated quantities.

Five glassy GeO$_2$ models containing 3240 atoms (Ge 1080 and O 2160 atoms, respectively) were also prepared using the RMC++ code \cite{evrard2005}. The initial structure was generated by randomly arranging atoms into cubic cells under periodic boundary conditions by hard sphere Monte Carlo; the number density was set to that measure by experiment at ambient conditions. The RMC modeling was performed based on structure factors from the neutron diffraction experiment with isotopic substitution \cite{salmon2002}. 
Coordination constraints around Ge atoms are used to avoid zero-to-three-fold coordinated Ge, which are the same constraints used in our previous work \cite{matsutani2022a} to avoid unphysical local structures. In this work, we additionally introduced coordination constraints around O atoms, which are used to prevent zero- and singly-coordinated O atoms.

\subsection{Ring analysis}
Here, we briefly describe the conventional ring definitions, which are used widely in the characterization of glass structures beyond short-range order. The first ring definition which mainly targeted SiO$_2$ glass was proposed by King in 1967 \cite{king1967a}; she defined a ring as the shortest path between two of the nearest neighbors of a given atom. This definition stems from the network of tetrahedral units. Using the same shortest path algorithms, Guttman proposed a different ring definition for silica \cite{guttman1990}. This defines a ring as the shortest path that comes back to a given atom from one of its nearest neighbors. It is known that sometimes there are large rings that are detected by King's definition but not by Guttman's definition. Guttman's rings can be seen as a subset of King's rings from this point of view.
Another definition, known as primitive ring \cite{yuan2002}, defines a ring as one that cannot be broken down into two smaller rings. 
For the characterization of each GeO$_2$ glass structure, we chose the above three ring definitions implemented by SOVA software \cite{sova}.

The ring analysis has been widely used to characterize glass structures, but the analysis ignores differences in the ring shape and length. Recently, ring shape characterization has been proposed \cite{salmon2023, shiga2023, kobayashi2023}, and in this study, we used the method proposed by Shiga \cite{shiga2023}, which is briefly described below. 
The method focuses on two parameters called roundness and roughness, both of which are easily calculated based on the eigenvectors and eigenvalue of the variance-covariance matrix of the atomic configurations composing the ring. The first eigenvector indicates the direction of the largest variance of those atom configurations. The second eigenvector represents the maximum variance perpendicular to this vector. In addition, the third eigenvector implies the maximum variance perpendicular to these vectors. Therefore, the shape of the ring can be measured by the proportion of these vectors.

\subsection{Persistent homology}
In addition to ring analysis, persistent homology (PH), which is implemented in the Homcloud software package \cite{obayashi2022}, is used to characterize the geometrical features within our models. 
PH analysis, based on the mathematical subject of algebraic topology, is a recently proposed tool for analyzing \textit{n}--dimensional holes in the given data set, such as atomic configurations generated by molecular dynamics and reverse Monte Carlo analysis. 
Theoretical details and applications of persistent homology can be found in the literature \cite{hiraoka2016, obayashi2022}; in this section, we will briefly outline how it is calculated. Simply put, this method identifies holes in the data and describes the structure of one-dimensional holes, namely rings. Firstly, a sphere with a common radius is placed at the coordinate of each atom regardless of the atomic species. Secondly, these radii are gradually increased, causing the spheres to overlap and generate new edges, gradually forming a ring. The time at which a ring is generated is recorded as its birth, corresponding to the horizontal axis in a persistence diagram (PD). As the radii continue to increase, the spheres further continue to overlap and extinguish the ring, which is recorded as its death, corresponding to the vertical axis in the PD. This process can be applied to the atomic configurations, allowing us to obtain a persistence diagram of the entire system, representing the rings and cavities embedded in the atomic configuration. We note that PD diagrams are often constructed by using the square of the radii according to conventions in computational geometry and TDA \cite{obayashi2022}, and we also use this convention in presenting our data.
This work will only focus on the one-dimensional holes which corresponds to rings.

\section{Results \& discussion}
\subsection{Validation of neural network potential}
\begin{figure}
    \centering
    \includegraphics{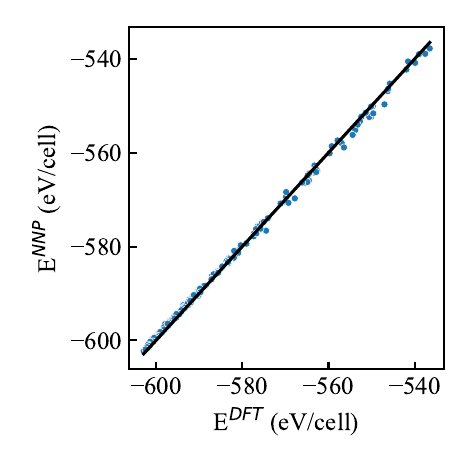}
    \caption{Comparison between ab initio calculation and NNP on the total energies for the test dataset.}
    \label{fig:val_E}
\end{figure}

First, we investigated the quality of the generated NNP using the test dataset. We ran AIMD simulations of the quenching process from 3000 to 300 K on a structure (96 atoms) independent of the training data, extracted the instantaneous structure every 10 ps, performed calculations to improve the accuracy of the AIMD calculations as well as the training data set, and obtained 200 structures as a test data set.
In Figure \ref{fig:val_E}, the energies predicted by the NNP are plotted against the AIMD energies of the test set. We can see that the predicted energies of our NNP are in good agreement with the reference AIMD energies, and the RMSE is 7.923 meV/atom, which is within acceptable values comparable to the literature \cite{lee2020}. Thus, the NNP have the capability to correctly describe interactions between atoms from liquid to glass in GeO$_2$. 

\subsection{Two-body correlations}
\begin{figure}
    \centering
    \includegraphics[width=\columnwidth]{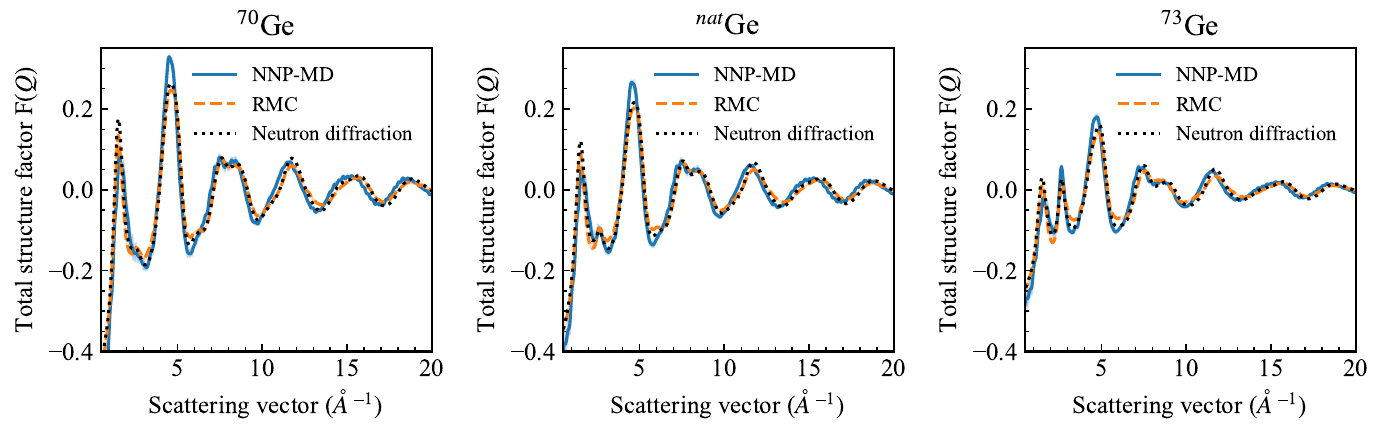}
    \caption{The comparison between NNP-MD, RMC, and neutron diffraction experiment \cite{salmon2007b} of the total structure factors $^{70}F(Q)$, $^{nat}F(Q)$, and $^{73}F(Q)$ for GeO$_2$ glass. The shaded regions indicate a \SI{95}{\percent} confidence interval.}
    \label{fig:tfk}
\end{figure}

\begin{figure}
    \centering
    \includegraphics{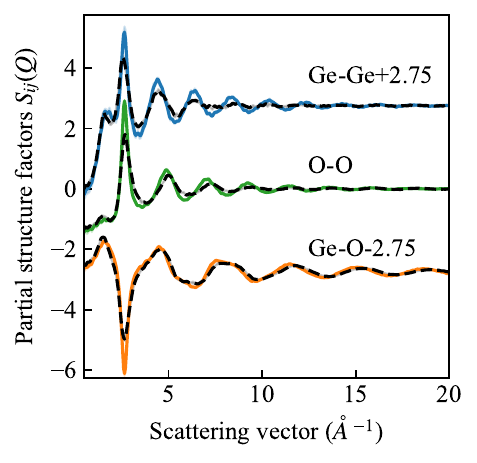}
    \caption{The comparison of the Faber-Ziman structure factors with NNP--MD (solid lines) and RMC (dashed lines). The shaded regions imply a \SI{95}{\percent} confidence interval.}
    \label{fig:psq}
\end{figure}

In the previous section, we confirmed that the generated NNP is numerically well-trained. This NNP was used to perform melt-quench simulations in the 3240-atom model to obtain glass structures. The structure factor is compared to RMC results in Fig \ref{fig:tfk}. The structure factors calculated from 3 independent NNP-MD trajectories are almost identical as evident from the very thin \SI{95}{\percent} confidence interval.
In general, good agreement between the NNP-MD and RMC models is observed in the heights and positions of prominent peaks in the structure factor. However, upon closer observation, it is seen that the RMC models exhibit a slightly higher FSDP which is closer to experiment than the NNP-MD models. On the other hand, the region around the so-called principal peak at $\sim$ \SI{2.64}{\angstrom}$^{-1}$ is reproduced better by NNP-MD. There are subtle differences in other peaks as well, but in general, the experimental structure factor, RMC models, and NNP-models are in good agreement with each other. The same can be said for the partial structure factors from RMC and NNP modeling, although the peaks are more slightly more diffuse for the RMC models (Fig.~\ref{fig:psq}).

\subsection{Local coordination environment}
The coordination numbers around Ge and O atoms are tabulated in Table \ref{tab:cn_change}. 
From this table, we can see that all of our models form a prototypical network structure in which tetrahedral units are cross-linked by O atoms. However, the NNP models exhibit a stronger inclination towards 4-coordination around Ge atoms and 2-coordination around O atoms compared to the RMC models. 

\begin{figure}
    \centering
    \includegraphics[width=\columnwidth]{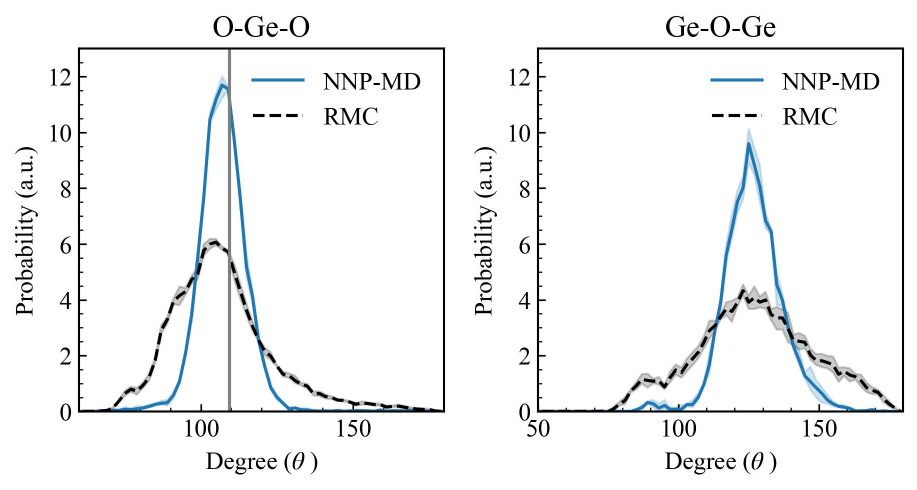}
    \caption{The bond angle distributions of structure models from reverse Monte Carlo analysis and NNP--MD. The shaded regions indicate a \SI{95}{\percent} confidence interval. The gray vertical lines indicate ideal intra-tetrahedral angles.}
    \label{fig:bads}
\end{figure}

\begin{table}[hbtp]
    \caption{The coordination environment around each atom.}
    \label{tab:cn_change}
    \centering
    \scalebox{1}{
    \begin{tabular}{c|cccc|cccccc}
        \hline
         & \multicolumn{4}{c|}{Around Ge} & \multicolumn{6}{c}{Around O}\\
         \hline
         & 4--fold & 5--fold & 6--fold & average & 1--fold & 2--fold & 3--fold & 4--fold & 5--fold & average\\
        \hline \hline
        GGA--PBE & 98.210 & 1.790 & 0.000 & 4.018 & 0.000 & 99.090 & 0.895 & 0.015 & 0.000 & 2.009\\
        RMC & 96.444 & 3.556 & 0.000 & 4.036 & 0.000 & 98.222 & 1.778 & 0.000 & 0.000 & 2.018\\
        \hline
    \end{tabular}
    }
\end{table}
The bond angle distribution, which is an indicator of three-body correlations, is widely used to describe geometric features in glass structures beyond two-body correlations. Here, we calculated the intra- and inter-polyhedral bond angle distributions (O--Ge--O and Ge--O--Ge) for each structure model (Fig. \ref{fig:bads}). The peak in the intra-polyhedral (O--Ge--O) bond angle distribution for both NNP and RMC models at around \SI{109.5}{\degree} indicates the existence of tetrahedral units. However, the O--Ge--O distribution for the RMC model has a much broader peak than that of NNP, indicating the higher abundance of distorted tetrahedral units. 
The interpolyhedral bond angle distributions (Ge--O--Ge) also show broader peaks in the RMC models, but the peak position at $\sim$ 125$^\circ$ is roughly in agreement between the RMC and NNP models. 
The shoulder around \SIrange{90}{100}{\degree} is present in both models as well, which correspond to chemical bonds between edge-sharing units or distorted units. This is in line with the general trend of the RMC method to result in broader peaks compared to AIMD \cite{akola2013}.

\subsection{Rings}
\begin{figure}
    \centering
    \includegraphics[width=\columnwidth]{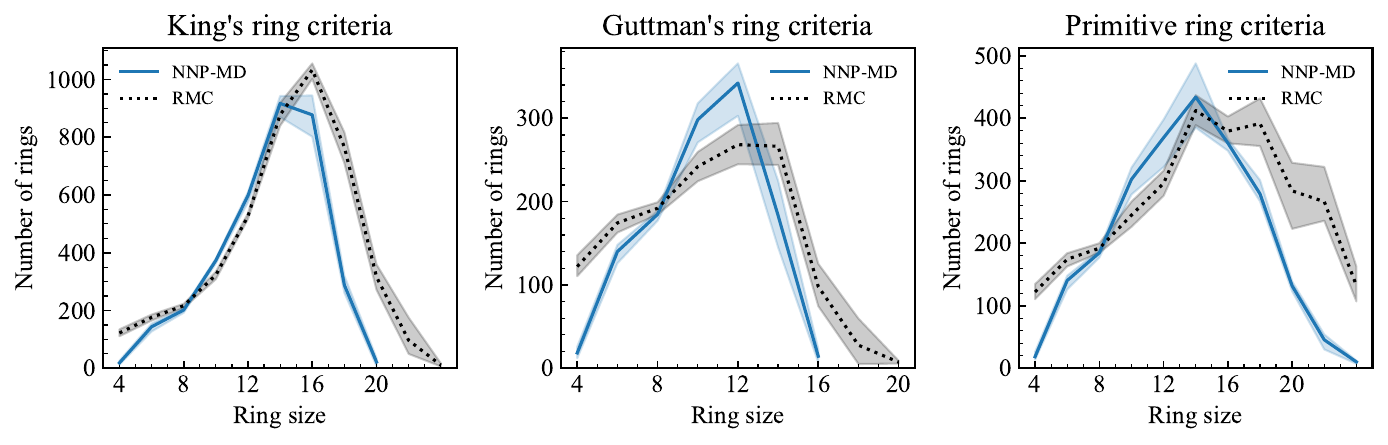}
    \caption{The ring size distributions of NNP models and RMC models. The number of each ring is normalized by the total number of atoms. The shaded regions imply a \SI{95}{\percent} confidence interval.} 
    \label{fig:ring_dist}
\end{figure}

The ring size distribution analysis has been widely used to characterize glass structure beyond two-- or three--body correlations. 
This study employed King's, Guttman's, and primitive ring criteria to characterize intermediate--range order in GeO$_2$ glass to clarify the structural difference between RMC and NNP--MD structural models. 
Figure \ref{fig:ring_dist} shows the average ring size distributions based on three ring criteria of the NNP and RMC structural models. All ring size distributions of RMC are broader than NNP-MD in this study; the ratio of 4- and 6-membered rings as well as the ratio of 16-, 14-, and 18-membered rings according to King's, Guttman's, and primitive ring criteria, respectively are larger than NNP-MD, and the peaks in the ring size distributions are also shifted to slightly larger rings in the RMC model. The behavior described above show the same trends as in the interpolyhedral bond angles (Figure \ref{fig:bads}). The peak around \SI{90}{\degree} corresponds to the presence of edge-sharing units, corresponding to 4-membered rings in the ring size distribution analysis, and the broad shoulder around \SI{150}{\degree} to the larger rings that we do not observe in the NNP-MD results. 
We note that a previous RMC analysis \cite{kohara2005} reported even larger fractions of 6- and 7-membered rings (they counted only Ge atoms in a ring structure, so 6- and 7-membered rings in previous work mean 12- and 14-membered rings).  These differences may be due to the different coordination constraints, as they only allow the existence of 4-fold Ge atoms and 2-fold O atoms in the previous RMC modeling, i.e., there are only corner-sharing units in their structural model. In our RMC models, however, the existence of edge-sharing units and 5-fold Ge and 3-fold O atoms is allowed based on the observation that these structural units exist in AIMD and NNP-MD results.

\begin{figure}
    \centering
    \includegraphics[width=\columnwidth]{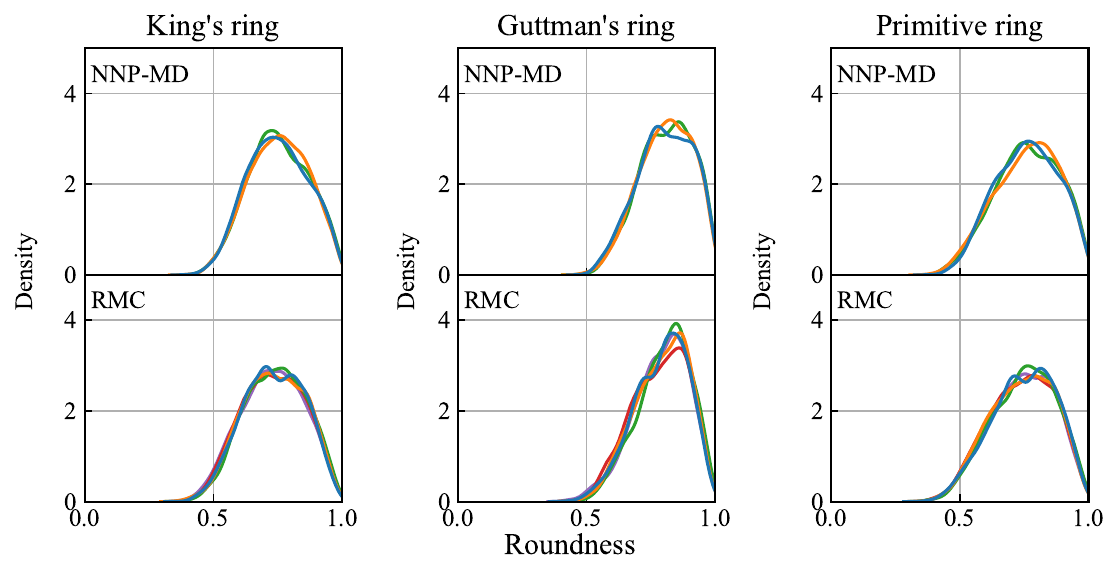}
    \caption{Kernel density estimate plots of the roundness compared between NNP--MD and RMC using three ring definitions. Here, we present results for each model realization, unlike previous figures where averages of independently generated models are presented.}
    \label{fig:roundness}
\end{figure}

\begin{figure}
    \centering
    \includegraphics[width=\columnwidth]{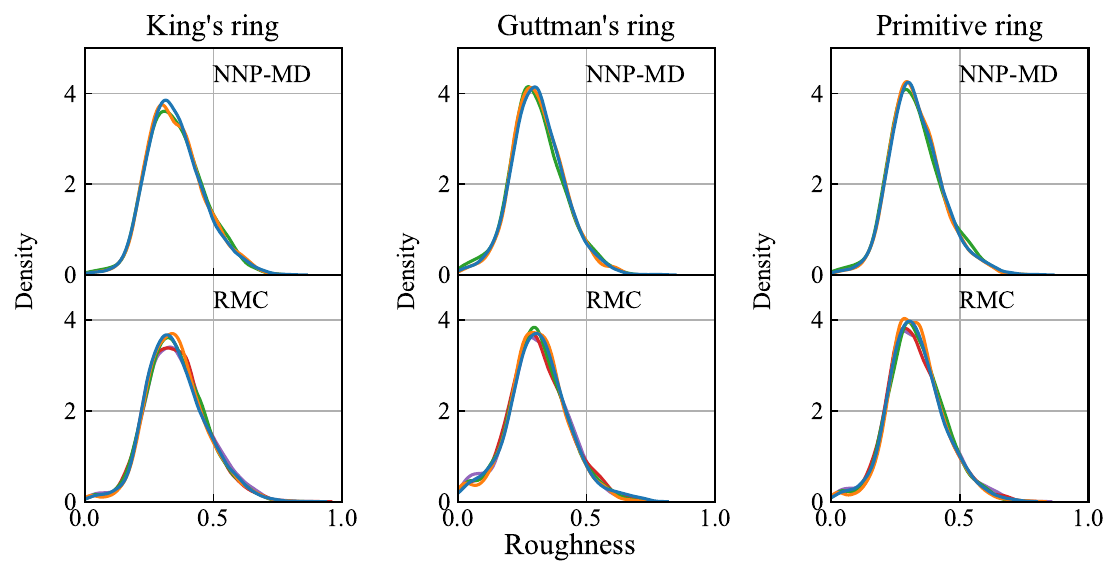}
    \caption{Kernel density estimate plots of the roughness compared between NNP--MD and RMC using three ring definitions.}
    \label{fig:roughness}
\end{figure}

\begin{figure}
    \centering
    \includegraphics[width=0.5\linewidth]{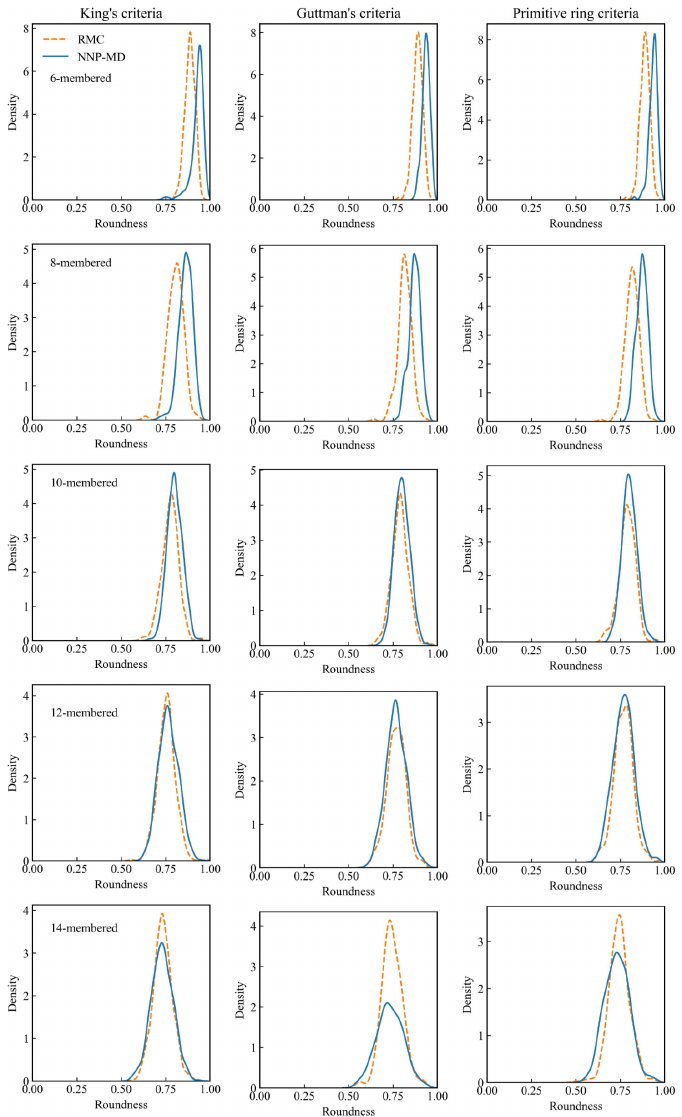}
    \caption{Comparison between RMC (dashed lines) and NNP-MD (solid lines) results of roundness parameter results for each ring size (6- to 14-membered rings)}
    \label{fig:round_each}
\end{figure}

\begin{figure}
    \centering
    \includegraphics[width=0.5\linewidth]{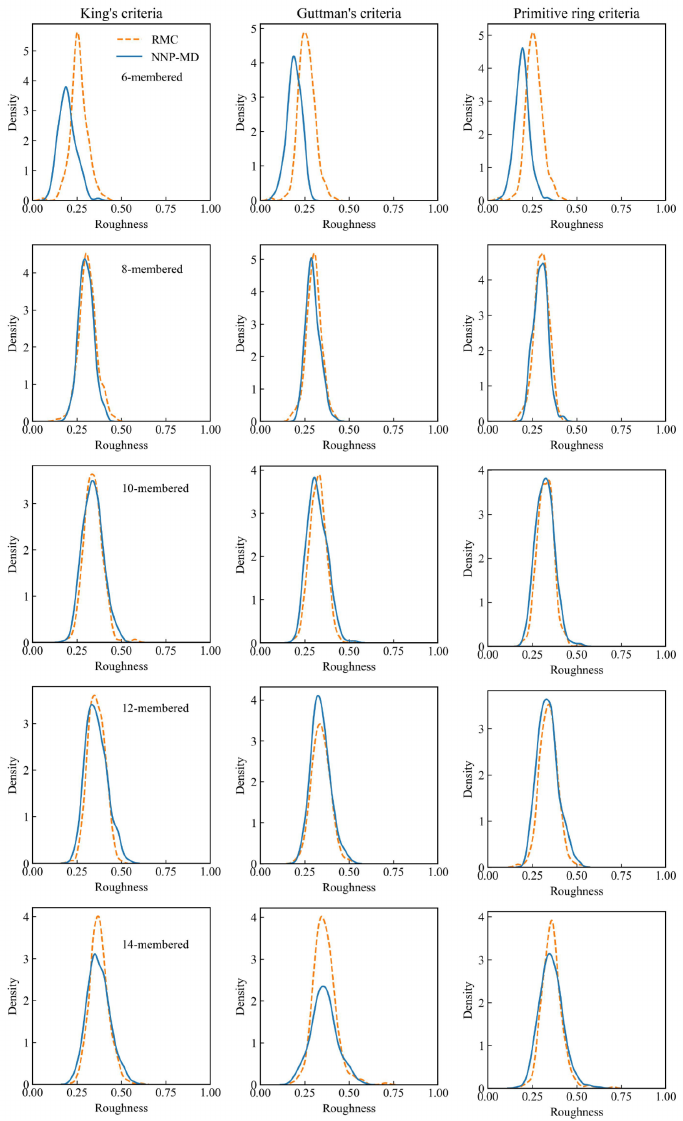}
    \caption{Comparison between RMC (dashed lines) and NNP-MD (solid lines) results of roughness parameter results for each ring size (6- to 14-membered rings)}
    \label{fig:rough_each}
\end{figure}

Figures \ref{fig:roundness} and \ref{fig:roughness} show the ring shape characterization of our NNP and RMC models based on the three ring definitions.
Ring shape distributions for each ring definition are essentially the same for the roundness and roughness parameters.
For further analysis, we calculated the roundness and roughness for each ring size (Figures \ref{fig:round_each} and \ref{fig:rough_each}). The results showed no significant differences, except for the smaller rings (6- and 8-membered rings). The RMC structures exhibit lower roundness values for the 6- and 8-membered rings compared to the NNP structures. Conversely, the RMC structures show higher roughness values for the 6-membered rings. This behavior may be attributed to the RMC's tendency to produce a more random structure.

\subsection{Persistent homology}
\begin{figure}
    \centering
    \includegraphics[width=\columnwidth]{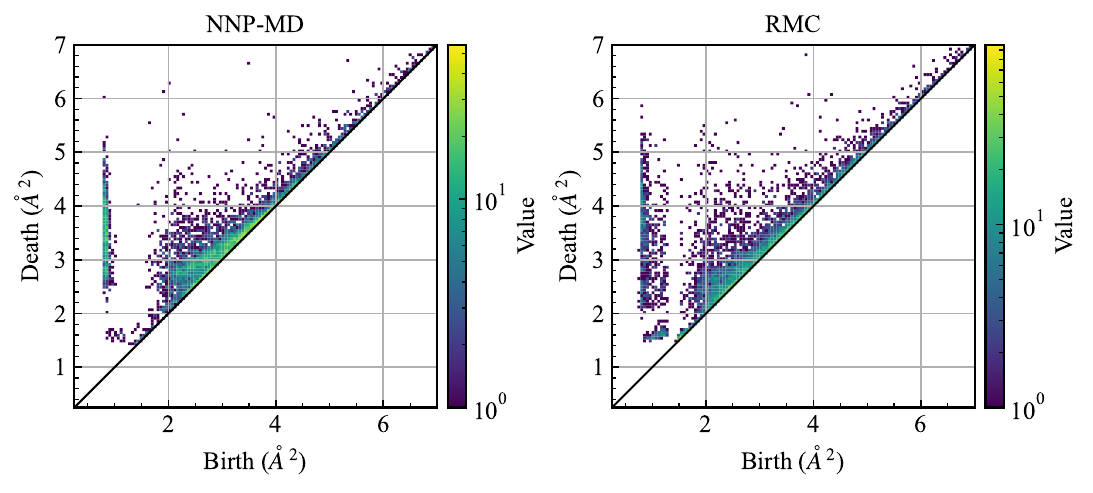}
    \caption{The 1--dimensional persistence diagrams for each structure model, respectively.}
    \label{fig:PD1}
\end{figure}

Figure \ref{fig:PD1} compares the persistence diagrams obtained by topological data analysis based on RMC and NNP models. 
Those persistence diagrams imply various ring distributions and the geometrical difference between our models. We can see that the island of birth pairs around \SI{0.8}{\angstrom} is broadened towards larger birth values in the RMC model compared to NNP models in line with the general tendency of RMC to favor more random structures as also seen in the bond angle and ring distributions above.
On the other hand, NNP tend to strongly restrict covalent bond lengths, as implied by the small variation in the birth value of these islands.

\begin{figure}
    \centering
    \includegraphics[width=\columnwidth]{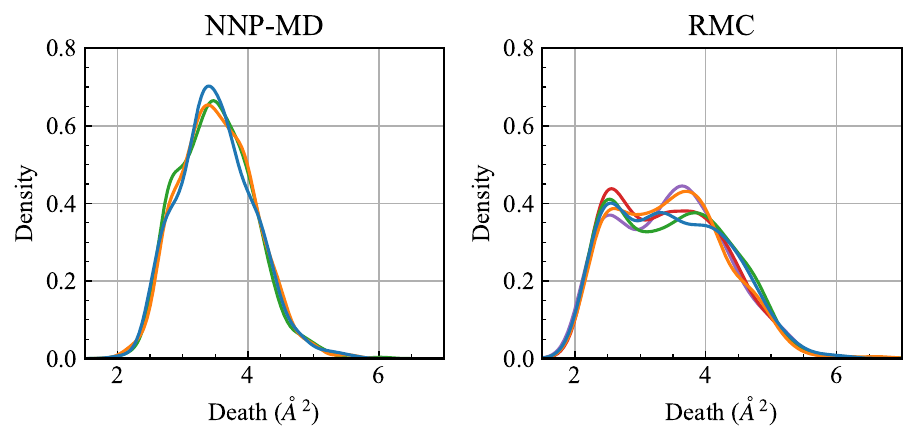}
    \caption{Kernel density estimate plots of the distribution of death scales in restricted areas is described in the main text compared between NNP-MD and RMC results.}
    \label{fig:death}
\end{figure}

Next, to further explore the differences between our models, the number of pairs present on the island over a death range of \SIrange{2}{7}{\angstrom^{2}} and births range of \SIrange{0.8}{1.266}{\angstrom^2}, which corresponds to the cutoff radius of the real space Ge--O bond length, was counted, yielding the results in Fig. \ref{fig:death}. 
The RMC death distribution appears to be a bimodal distribution, with the first peak located around \SI{2.5}{\angstrom^{2}} and the second peak located around \SI{3.75}{\angstrom^{2}}. 
However, the NNP models only have a single sharp peak which is located around \SI{3.5}{\angstrom^{2}}.
These behaviors appear similar to the results of traditional ring analysis, even though no ring definition is used, but unlike traditional ring distributions, differences can be seen on the length scales.

\section{Summary}
We constructed neural network potentials and performed extensive NNP--MD simulations to explore the structural features in glass structure and compared them with RMC analysis.
The goal was to provide the real space image of intermediate-range order in the GeO$_2$ glass and evaluate the uncertainty of the RMC analysis based on simple constraints and neutron diffraction data, and our findings are summarized as follows:
\begin{itemize}
    \item The structural models obtained from the NNP are in good agreement with the experimental results and the RMC analysis in terms of two-body correlations. 
    \item Bond angle distributions are highly broadened in the RMC model compared to the NNP-MD model; this is in line with the well-known tendency of RMC to lead to more random structures. 
    \item The intermediate range order as investigated by ring and topological data analysis leads to a sharper distribution in the NNP models and the underlying density functional method places stronger and probably more accurate restrictions on the intermediate range order in the glass structure.
\end{itemize}
In other words, the well-known shortcoming of relying only on two-body correlations for real-space modeling also affects the resulting intermediate-range order. This study demonstrates the promise of structural modeling with neural network potentials molecular dynamics for overcoming this issue.

\section*{Acknowledgments}
We would like to thank Philip Salmon for providing us with his neutron diffraction data, and Professor Motoki Shiga and Associate Professor Kazuto Akagi for their valuable advice on ring analysis and persistent homology.
This research was financially supported by Japan Science and Technology Agency FOREST Program (Grant Number JPMJFR2037 Japan), supported by Japan Science and Technology Agency, the establishment of University fellowships towards the creation of science technology innovation, Grant Number JPMJFS2104 and also supported by the Japan Soceity for Promotion of Science (JSPS) KAKENHI Grant Number 20H02430.
Computations were performed at the Super computer Center, ISSP, the Univ. of Tokyo.

\section*{Author declarations}
\subsection*{Conflict of Interest}
The authors have no conflicts to disclose.

\section*{Data availabilty}
The datasets generated during this research are openly available from the ISSP Data Repository at https://datarepo.mdcl.issp.u-tokyo.ac.jp


%
%

%


\bibliography{GeO2_ambient_NNP_RMC}

\end{document}